\documentclass[prl,twocolumn,superscriptaddress,amsmath]{revtex4}
\usepackage{graphicx}
\usepackage{float}

\begin{document}

\def\ket#1{|#1\rangle} 
\def\bra#1{\langle#1|}
\def\av#1{\langle#1\rangle}

\def\r{\color{red}}
\def\b{\color{black}}

\title{Passive, deterministic photonic CPHASE gate via two-level systems}

\author{William Konyk}
\affiliation{Department of Physics, University of Arkansas, Fayetteville, AR 72701\\}
\author{Julio Gea-Banacloche}
\affiliation{Department of Physics, University of Arkansas, Fayetteville, AR 72701\\}

\date{\today}

\begin{abstract}
We show that an array of identical two level systems coupled losslessly to a one dimensional waveguide is able to realize a high fidelity conditional phase shift useful for quantum logic. We propose two arrangements of emitters (one that relies on direct interactions between the emitters, and one that does not), and describe possible physical realizations and limitations.
\end{abstract}
\maketitle

Photons are one of the best carriers of quantum information, as they interact only weakly with the environment, and photon processes are fast and efficient. Despite these advantages, two main issues have prevented all-optical quantum computing from being realized as effectively as computing with ions, trapped atoms, or superconducting qubits: true on-demand single photon sources are difficult to create and photon-photon interactions are challenging to engineer. In this paper we propose a realistic means of circumventing this second challenge and demonstrate how two level emitters (TLEs) coupled to a one-dimensional waveguide can, in principle, implement a passive, near-deterministic conditional phase (CPHASE) gate between two photons. 

There have been many proposals to overcome this second challenge. Many schemes of all-optical QIP (such as \cite{LOQC}) only work probabilistically, requiring a substantial overhead. Other proposals require active elements or control of an optical medium \cite{ActGaus,Duan,Hacker} but this typically makes them difficult to scale up to large computations. Schemes based on temporary storage of one of the photons are either complicated by the need for several steps \cite{ZhengWQED}, or require quantum memories for intermediate storage \cite{KIN}.  Ideally, what one would want is to simply send two photons flying through a nonlinear medium and have them exit with a useful, conditional phase, as originally proposed in \cite{chuang}.  However, existing, conventional media do not exhibit sufficiently strong optical nonlinearities, and may be too ``noisy'' for quantum information processing applications \cite{Shapiro}, while spectral entanglement has also been identified as a potential issue in these schemes \cite{GeaPhase}. 

A recent important step in this direction was provided by Brod and Combes in \cite{brod1,brod2}, where they showed that an array of two level systems can, in principle, allow two photons to conditionally interact without becoming spectrally entangled. Central to their scheme is the idea that the photons must propagate in opposite directions through the array, requiring either a perfect chiral coupling between the emitters and the waveguide, or the use of optical circulators at every step.  Additionally, their proposal required some way to have the TLEs interact while keeping the photons physically separate.  

Here we show that an array of two-level atoms coupled to an ordinary (non-chiral) waveguide can, in principle, perform the CPHASE operation between two counterpropagating, single-photon wavepackets. 
Our scheme relies on the existence of transmission resonances in the interaction of a single photon with a pair of TLEs, which we pointed out in \cite{us2} and which have been interpreted as Fano resonances by other authors (\cite{agarwal}; see also \cite{baranger,law,liao}).  We showed in \cite{us2} that, under the right conditions, these transmission windows persist in the nonlinear regime where two counterpropagating photons interact with the pair of TLEs, so the photons are transmitted with near-unit probability, but they pick up a nontrivial phase as they do so.  

Ideally, a conditional phase gate would impart a phase $\phi_1$ to the states $\ket 0\ket 1$ and $\ket 1\ket 0$ (where 1 and 0 refer to the presence or absence of a photon) and a phase $\phi_2$ to the state $\ket 1\ket 1$.  As long as $\Phi=\phi_2-2\phi_1\neq 0$, this gate, together with single photon gates, would enable universal quantum computation. In practice, we desire $\Phi$ to be as large as possible while maintaining a high fidelity, as discussed below.  In our scheme, as we show here, $\Phi=\pi$ is possible for an array of pairwise interacting atoms, and $\Phi = \pi/2$ for a system without interactions. 

Regarding the issue of fidelity, we note that (in the same way as \cite{brod1}), our gate will distort the spectrum of a single photon. As pointed out by Brod and Combes, this can be overcome by ensuring that at every step in the computation all photons are distorted in the same way. We then define the fidelity of the two-photon operation relative to the product state of two independent single-photon transmissions through the array of $N$ pairs of scatterers.  Formally, we write
\begin{equation}
    \sqrt{\mathcal{F}}e^{i\Phi}=\bra{\text{Target}} \psi_{\text{Scattered}}\rangle \label{fiddef}
\end{equation}
where the target state has the form $\ket{\text{Target}}= \int d\omega_1 d \omega_2 t_N(\omega_1) t_N(\omega_2) f_0(\omega_1)f_0(\omega_2) \hat a^{\dagger}_{\omega_1} \hat b^{\dagger}_{\omega_2}\ket{0}$, and $t_N(\omega)$ is an appropriate single photon transmission coefficient for the array of $N$ sites.
  
The first scheme that we consider utilizes the transmission window presented in Fig. 4 of \cite{us2} that is created when two identical emitters can directly exchange a quanta of energy. Specifically, we postulate an interaction between neighboring TLEs of the form 
\begin{equation}
H_I = \hbar\Delta\bigl(\ket{eg}\bra{ge}+\ket{ge}\bra{eg}\bigr)
\end{equation}
(where $\ket g$ and $\ket e$ denote the ground and excited states, respectively).  This is a F\"orster-type interaction that may arise naturally between quantum dots \cite{forster}, and it has also been used by other authors to model the dipole-dipole interaction between neighboring two-level atoms \cite{deutsch}.  As we showed in  \cite{us2}, when the condition $\sin(2\pi d/\lambda)=-\Delta/g^2 = -1$ is satisfied (where $d$ is the distance between the emitters, $\lambda$ the central wavelength of the photons, and $g$ the coupling between the emitters and the waveguide mode), and the incoming photons are on resonance, two counterpropagating photons will be perfectly transmitted through the pair of emitters, while their spectra are modified according to the single-pair scattering matrix
\begin{align}
    &S_1(\omega_1,\omega_2,\nu_a,\nu_b)= t_1(\omega_1)\delta(\omega_1-\nu_a) t_1(\omega_2)\delta(\omega_2-\nu_b)\cr
    &\quad -\frac{2 \Gamma}{\pi} \Bigg(\frac{1}{\Gamma-i\omega_1}+\frac{1}{\Gamma-i\omega_2}\Bigg)\frac{\delta(\omega_1+\omega_2-\nu_a-\nu_b)}{(\Gamma-i\nu_a)(\Gamma-i\nu_b)} \label{IntSsing}
\end{align}
where henceforth $\Gamma\equiv g^2$, and $t_1(\omega) = -(\Gamma+i\omega)/(\Gamma-i\omega)$. This scattering matrix turns out to be identical to the single site scattering matrix of Eq.~(29) in \cite{brod2} when the interaction proposed there has an infinite strength (i.e. $\chi\rightarrow \infty$), which is also the limit in which their gate operates optimally.  Here, we see how this result may arise from a finite-strength, realistic interaction between TLEs, and, because of the perfect transmission property, no special effort is required to ensure a chiral coupling.  

\begin{figure}
\centering
\includegraphics[width=\columnwidth]{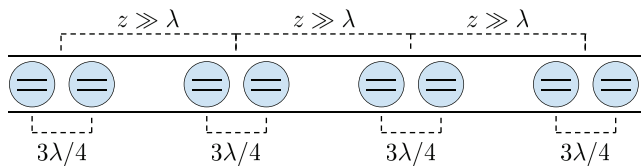}
\caption{ \label{IntGate}  A schematic for the construction of a gate using pairs of emitters with a direct energy-exchange interaction; TLEs are arranged in closely-spaced pairs, and successive pairs are placed far enough apart so that the interaction between pairs is negligible.}
\end{figure}

To build the equivalent of the Brod-Combes gate, one could then arrange pairs of emitters as in Fig. \ref{IntGate}, where the distance between emitters in a pair is $3/4$ of a wavelength, and each pair is separated by many wavelengths from its nearest neighbor, to ensure that the interaction between emitters in different pairs is negligible. A possible physical realization of such an arrangement would consist of placing closely-spaced quantum dots or pairs of superconducting circuits connected by a wire within a photonic crystal waveguide, as this would ensure strong coupling to the guided modes and provide a realistic means of ensuring direct energy transfer between emitters.  

\begin{figure}
\centering
\includegraphics[width=\columnwidth]{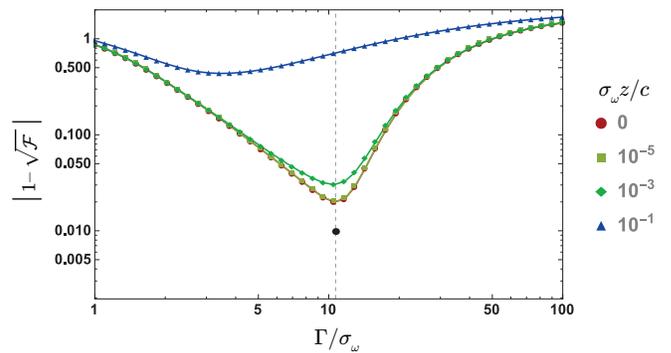}
\caption{ \label{EmitSep}  Dependance of the phase and fidelity of the gate operation on the separation between neighboring sites, $z$, for $N=12$ pairs of emitters.  This was calculated for an unentangled initial state with a Gaussian profile of width (standard deviation) $\sigma_\omega$.  $c$ is the speed of light in the waveguide. The dashed line shows the value of $\Gamma/\sigma_\omega$ for which maximum fidelity was predicted in \cite{brod1}.  The black dot shows the prediction of \cite{brod1} for the gate fidelity, a somewhat different measure from the one we have adopted here.}
\end{figure}

Since the photons are transmitted through each pair with unit probability, they can only interact at one pair; after this, they will just continue to transmit through the array in opposite directions. This limits the number of scattering channels, and for $N$ pairs of emitters, each with center located at $z_j$, the scattering matrix is
\begin{align}
    S(\omega_1,\omega_2,\nu_a,\nu_b)= t_1(\omega_1)^N\delta(\omega_1-\nu_a) t_1(\omega_2)^N\delta(\omega_2-\nu_b)\cr
    -\frac{2 \Gamma}{\pi}\sum_{j=1}^N t_1(\omega_1)^{N-j} t_1(\omega_2)^{j-1} t_1(\nu_a)^{j-1} t_1(\nu_b)^{N-j} \cr
    \times \Bigg(\frac{e^{2 i (\nu_a-\omega_1)z_j/c}}{\Gamma-i\omega_1}+\frac{e^{-2 i (\nu_b-\omega_2)z_j/c}}{\Gamma-i\omega_2}\Bigg)\frac{\delta(\omega_1+\omega_2-\nu_a-\nu_b)}{(\Gamma-i\nu_a)(\Gamma-i\nu_b)} \label{IntSTot}
\end{align}
This result is identical to the $\chi \rightarrow \infty$ limit of the scattering matrix given in Eq.~(60) of \cite{brod2}, provided all $z_j=0$. By including the positions of the pairs, we are able to study the effect of emitter separation and placement in the fidelity of the gate, something that was absent in the analysis in \cite{brod1}. In Fig.~\ref{EmitSep} we show how, for $N=12$ sites, emitter separation modifies the fidelity of the operation.

We may understand Fig.~\ref{EmitSep} by extending an argument first presented in \cite{brod1}.  Suppose each photon spends a time of the order of $1/\Gamma$ interacting with each pair of TLEs, and a time $z/c$ in transit from one pair to the next.  We get a total time $N/\Gamma + Nz/c$, and we expect optimal coupling when the initial duration of the pulse, $1/\sigma_\omega$, is of this order of magnitude, that is, $N\sigma_\omega/\Gamma + N\sigma_\omega z/c \sim 1$.  In addition, we want $\sigma_\omega z/c \ll  \sigma_\omega/\Gamma$ to insure that the photons spend most of their time at the interaction sites, so they can meet at one of them, rather than ``passing by each other'' in transit.  Lastly, for reasons discussed at length in \cite{brod2}, we want to be in the strong adiabatic limit where $\Gamma \gg \sigma_\omega$.  The figure shows how all these requirements play against one another, for the relatively small number of sites $N=12$.

The particular parameter choice we have discussed so far, namely $\Delta=\Gamma$ and $d= 3\lambda/4$ in each pair, has the formal advantage of reducing exactly to the Brod-Combes gate, and of exhibiting a very broad transmission window, but it may be very difficult to satisfy in practice.   Alternative (although, in general, narrower) transmission windows can be found for other values of $\Delta$, and even for $\Delta =0$ (which means no direct interaction between the TLEs in a pair), provided a finite detuning is introduced.  For simplicity, therefore, we will consider the $\Delta =0$ case in the remainder of this paper, and show that in this limit also a usable conditional phase can  be obtained.   

For non-interacting TLEs, we found in \cite{us} that high transmission of counterpropagating photons  through a pair of TLEs will happen for a detuning $\delta$ and distance $d$ satisfying $\tan(2\pi d/\lambda)=-\delta/\Gamma$ (see Fig.~3 and Eq.~(24) of \cite{us2}, for the single-photon and Fig.~8 and Eq.~(49) for the two-photon case).  However, the narrow transmission window means that for a finite bandwidth pulse there will always be a small reflection probability, and the cumulative effect of this could completely ruin the performance of an array consisting of a large number of pairs. 

\begin{figure}[t]
\centering
\includegraphics[width=\columnwidth]{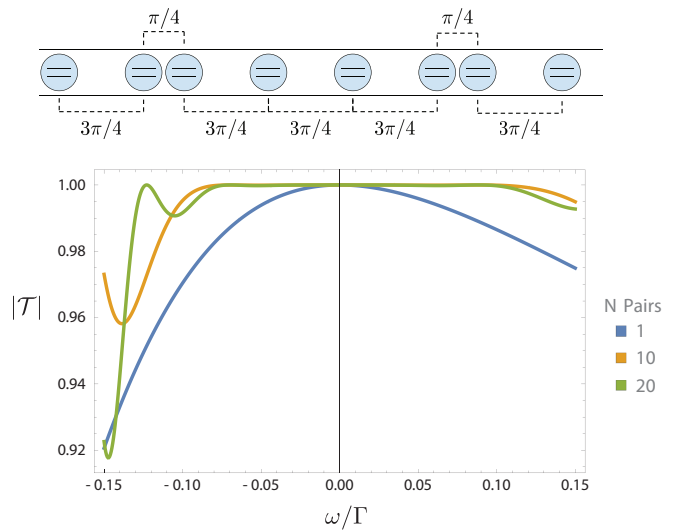}
\caption{ \label{NoIntSetup}  Top: A schematic for the construction of a gate using non-interacting emitters. The values refer to the phase difference that a photon experiences when traveling from one emitter to the next, and thus not necessarily to the physical distance between emitters. Bottom: The intensity transmission coefficient, $\mathcal{T}=|t(\omega)|^2$, as a function of dimensionless parameter $\omega/\Gamma$.}
\end{figure}

To address this problem, we have looked at ways to place the various pairs so as to maximize the overall transmission probability through the array.  For a single photon, this can be done by describing each pair of TLEs by its (frequency-dependent) reflection and transmission coefficients, and arranging the distance between pairs to maximize destructive interference of the reflected fields.  The result is a broader and flatter transmission curve as a function of frequency for the whole array, as seen in Figure \ref{NoIntSetup}, so a Gaussian pulse can be nearly perfectly transmitted.  The particular arrangement shown in the top diagram in Fig.~\ref{NoIntSetup} was derived by analytically maximizing transmission between two pairs, then four pairs, then eight pairs, up to thirty two pairs. It is specifically chosen for the case $\delta =\Gamma$ (a convenient choice, as we show below). 

Under these conditions, we can approximate the effect of a single pair of atoms on a single incident photon as just the (frequency-dependent) phase factor $t_1(\omega) =  \exp[-2i\phi-2i\omega/\delta]$, where $\phi\equiv 2\pi d/\lambda$, in the adiabatic regime where both $\delta$ and $\Gamma$ are much greater than $\sigma_\omega$.  In this same regime, we can derive from Eq.~(46) of \cite{us} the following two-photon scattering matrix for a single site (to be compared to Eq.~(\ref{IntSTot})): 
\begin{align}
    &S_1(\omega_1,\omega_2,\nu_a,\nu_b)= t_1(\omega_1)\delta(\omega_1-\nu_a) t_1(\omega_2)\delta(\omega_2-\nu_b)\cr
&\qquad-{\cos^2\phi\over\pi} e^{-2i\phi} \delta(\omega_1+\omega_2-\nu_a-\nu_b) \times\cr   
    \biggl[&\Gamma_+^2\left({1\over \Gamma_+ -i\omega_1}+ {1\over \Gamma_+ -i\omega_2}\right) {1\over (\Gamma_+ -i\nu_a)(\Gamma_+ -i\nu_b)} \cr
&+\Gamma_-^2\left({1\over \Gamma_- -i\omega_1}+ {1\over \Gamma_- -i\omega_2}\right) {1\over (\Gamma_- -i\nu_a)(\Gamma_- -i\nu_b)} \biggr] 
\end{align}
where $\Gamma_{\pm} = \Gamma e^{i\phi}(1\pm\cos\phi)/\cos\phi$.  When this result is used to form the $N$-site scattering matrix (as in Eq.~(\ref{IntSTot})), the same treatment as in \cite{brod2} shows that in the adiabatic- and large-$N$ limit one should get a product state with a usable phase of $\Phi = -2\phi$, which is equal to $\pi/2$ for the choice $\delta=\Gamma$. 

To verify this prediction numerically, we have extended the time-domain solution for multi-photon pulses presented in \cite{us} and \cite{us2} so as to describe the scattering of two photons from an array of $N$ emitters at arbitrary positions. We present the full derivation in the supplementary material for this paper, as it is rather involved. The solution follows the same time-domain method and uses the ``Markovian approximation'' presented in \cite{us2}. Thus it holds as long as the separation between the farthest pairs of emitters is negligible compared to the spatial length of the photons. We have also verified this---that is to say, the validity of the Markovian approximation---using an exact calculation for single photons following the procedure given in \cite{zubairy1}.  We find that as long as the term $\sigma_\omega z/c\approx 10^{-3}$, with $z$ again being the separation between neighboring pairs, the exact and Markovian solutions only differ by around 1\% for up to $N=50$ pairs.

\begin{figure}[t]
\centering
\includegraphics[width=\columnwidth]{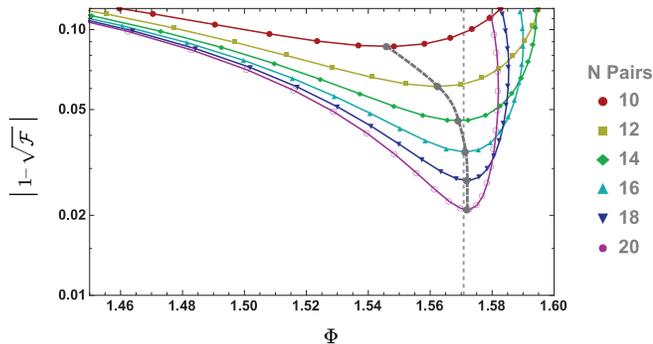}
\caption{ \label{NoIntFid}  The infidelity as a function of useful phase shift for the non-interacting gate design for various numbers of sites (increasing from top to bottom). The points of maximum fidelity have been connected by a dashed line, to exhibit the asymptotic behavior. }
\end{figure}

Fig.~\ref{NoIntFid} shows the results of this numerical calculation for the fidelity and phase defined as in Eq.~\ref{fiddef}. Clearly, as the number of pairs increases the infidelity goes down, and we find a usable phase $\Phi = \phi_2 -2\phi_1$ that approaches $\pi/2$.  The infidelity follows a power law well, and is given by $1-\sqrt{\mathcal{F}_{max}}=8.16 N^{-1.97}$, where $N$ is the number of pairs of TLEs. The points of maximum fidelity occur at $\Gamma/\sigma_\omega=3.67 N^{0.5536}$. This scales similarly to the design using interacting pairs, as there the infidelity follows $1-\sqrt{\mathcal{F}}=.988 N^{-1.57}$ and maximum fidelity occurs (in agreement with \cite{brod1}, provided $\Gamma=\gamma/2$ as we use a bidirectional waveguide) when $\Gamma/\sigma_\omega=1.42 N^{.818}$. 

We turn now to consider the sensitivity of this gate to imperfections in the experimental setup.  The design we considered earlier, using interacting atoms, had the advantage of a virtually unbounded transmission bandwidth, although this was achieved at the cost of having to meet the very precise requirement $\Delta = \Gamma$ (where $\Delta$ is an interaction strength, which cannot be adjusted simply by changing the separation between the atoms, since that needs to satisfy $\sin(2\pi d/\lambda) = -1$).  The non-interacting scheme is much more flexible in that regard, since one can always adjust the detuning $\delta$ to have $\delta = \Gamma$, but the price is a narrower transmission window.  We find, in particular, that an exponentially-decaying pulse, with a Lorentzian spectrum, has a significantly lower transmission probability through the arrangement in Figure \ref{NoIntSetup}: the maximum fidelity for $N=14$ pairs is only .699, compared with .955 for a Gaussian of the same duration (as measured by the standard deviation in time for the wavepacket, $\sigma_t$).  However, it is certainly possible (as has been shown, for example,  in \cite{Kolchin}) to modify the shape of a single photon so that it has a Gaussian profile. While such a process introduces losses, it would only need to be performed once (for each photon), at the beginning of a computation. 

One feature common to both designs is that they rely on a very particular arrangement of emitters. Realistic systems will contain errors, however, and in order to estimate the sensitivity of this system to the placement of the TLEs we have calculated the phase and fidelity of the output state of the non-interacting design for $N=14$ pairs (chosen because it has a high fidelity but is relatively quick to compute), including random errors in the location of the emitters.  

\begin{figure}
\centering
\includegraphics[width=\columnwidth]{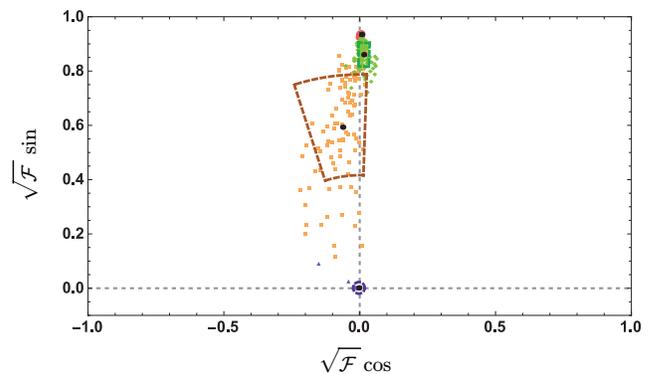}
\caption{ \label{PosErr}  A polar plot of the phase (angle) and fidelity (radius) of the two photon operation with random errors in the positions. The maximum allowable error has been scaled so that any individual atom may at most be offset by a percent of the wavelength given by, from top to bottom: $0.1\%$ (red); $0.1\%$ in the spacing between atoms in a pair and $10\% $ in the spacing between pairs (green); $1\%$ in both (orange); $10\% $ in both (purple). The dashed regions represent the standard deviation in the angle and the fidelity centered around the mean (black dot).}
\end{figure}

As can be seen in Fig. \ref{PosErr}, the system is very sensitive to the placement of the scatterers. An error of $0.1\%$ of the wavelength does not significantly affect the result, but a random error of $1\%$ of the wavelength both significantly decreases the phase and the fidelity. If the error in the placement of the scatterers has a maximum of $10\%$, the desired effect is gone. The most important parameter appears to be the spacing within an individual pair; when the error in the placement of emitters within a pair is on the order of $0.1\%$ of a wavelength, the spacing between pairs can differ by $10\%$ of a wavelength (green points in Fig. \ref{PosErr}), and the system still works roughly as expected.

Finally, our results do assume that the emitters only interact with the waveguide modes, which will probably be the most challenging requirement, since each photon needs to interact with a large number of scatterers.  We note, however, that coupling efficiencies in excess of $98\%$ have been reported for quantum dots in photonic crystal waveguides in \cite{lodahl}, so our system might not require a huge improvement over the current state of the art.

In conclusion, we have shown that a passive, deterministic, photon-photon CPHASE gate, along the lines first proposed by Brod and Combes \cite{brod1}, can be realized physically by an array of two level systems embedded in a waveguide, with no need for chiral coupling, optical circulators, or even direct interaction between the TLEs.  The nonlinear mechanism is, in essence, the competition of the two photons for the excitation of each atom, enhanced by an interference, or cavity-like, effect within each pair of atoms.  While a realistic gate based on these designs would inevitably contain errors, the fact that the fidelity of the desired operation can be made quite high, in principle, suggests that these systems could drastically reduce the overhead currently exhibited by schemes for fault-tolerant quantum computation with photons and probabilistic gates. This would be a significant step forward in constructing all-optical quantum computers. 

\bibliography{main_text}
\bibliographystyle{apsrev4-1}

\widetext
\clearpage
\begin{center}
\textbf{\large Supplemental Material: Passive, deterministic photonic CPHASE gate via two-level systems}
\end{center}
\setcounter{equation}{0}
\setcounter{figure}{0}
\setcounter{table}{0}
\setcounter{page}{1}
\makeatletter
\renewcommand{\theequation}{S\arabic{equation}}
\renewcommand{\thefigure}{S\arabic{figure}}

\section{Introduction}

We present here a number of results for the problem of two photons and $N_a$ two-level emitters (TLEs), in the case in which the TLEs do not directly interact with each other (in the notation of \cite{us2}, $\Delta=0$).  In Section 2 we show the formal solution to the problem, which we have used primarily for numerical calculations.  Section 3 shows in detail how the approximate form for the scattering matrix for a single pair of TLEs presented in the main text (Eq.~(5)) can be derived from the results we presented in \cite{us2}, and then iterated so as to yield an analytical approximation to the final scattering matrix of the two counterpropagating photons.  Section 4 describes the way we have used the single-pair results in \cite{us2} to optimize the separation between pairs in order to increase the width of the transmission resonance, as shown in Fig.~3 of the main text.

Throughout this supplement, we will use $N_a$ to denote the number of TLEs (``atoms''), and $N$ to denote the number of pairs, as we do in the main text, so $N = N_a/2$. 

\section{Full solution for two photons interacting with a one-dimensional array of $N_a$ scatterers}

This is an extension of the time-domain solution first presented in \cite{us}, and again in \cite{us2}, to deal with the case when one has $N_a$ identical two level emitters losslessly coupled to a one-dimensional waveguide at arbitrary positions, and interacting with a multimode two-photon field. The Hamiltonian for the system is 
\begin{equation}
H=\hbar g \sum_{j=1}^{N_a} \left[\hat \phi_{j}(t)e^{-i\delta t}\sigma_{j}^{\dagger}+\hat \phi_{j}^{\dagger}(t)e^{i\delta t}\sigma_{j}\right]
\label{ham}
\end{equation}
where the atomic operators $\sigma_{j}^{\dagger}$ and $\sigma_j$ raise or lower the j-th emitter, and the field operators are   
\begin{equation}
\hat \phi_{j}(t)=e^{i k_F z_j}\hat A\left(t-\frac{z_j}{c}\right)+e^{i k_F z_j}\hat B\left(t+\frac{z_j}{c}\right)
\end{equation}
The operators $\hat A(t)$ and $\hat B(t)$ annihilate a right- or left-propagating photon, respectively, at the time $t$ (see Eq.~(1) of \cite{us}). The operators $\hat \phi_{j}$ have the only non-vanishing commutator 
\begin{equation}
    \left[\hat \phi_j(t),\hat \phi_{k}^\dagger(t_1)\right]=e^{i k_F (z_j-z_k)}\delta\left(t-t_1-\frac{z_j-z_k}{c}\right)+e^{-i k_F (z_j-z_k)}\delta\left(t-t_1+\frac{z_j-z_k}{c}\right)
    \label{phicom}
\end{equation}
This commutator always appears in the context of terms like
\begin{equation}
    \int_{-\infty}^{t} dt_1 \left[\hat \phi_j(t),\hat \phi_{k}^\dagger(t_1)\right]e^{-i\delta t +i\delta t_1} \ket{\psi(t_1)}
\end{equation}
so, if $z_j>z_k$, only the first delta function in (\ref{phicom}) will contribute, and  if $z_j<z_k$, only the second one will. Either way, the phase factor can be written as $e^{i k_F |z_j-z_k|}$.  

At this point, we make the so-called ``Markovian approximation,'' which consists in neglecting the change in the state of the system over the time it takes for light to travel from one atom to another.  That is, we set $\ket{\psi(t\pm(z_j-z_i)/c)}\approx \ket{\psi(t)}$.  This is formally equivalent to replacing the commutator (\ref{phicom}) by
\begin{equation}
     \left[\hat \phi_j(t),\hat \phi_{k}^\dagger(t_1)\right]=2\theta_{j,k}\delta(t-t_1)
     \label{phicom2}
\end{equation}
where we have defined $\theta_{j,k}=e^{i k_F |z_j-z_k|}$ for compactness.

To solve for the final scattered photon state, we express the system's total state as
\begin{equation} \ket{\psi(t)}=\ket{\psi_{2,0}(t)}+\ket{\psi_{1,1}(t)}+\ket{\psi_{0,2}(t)}\end{equation}
where the first index refers to the number of photons and the second refers to the number of excited atoms. Each of these terms contains both atom and photon states. We use the same procedure introduced in \cite{us} where we normal-order all photon operators to arrive at a closed, finite series solution for any finite initial photon number (in this case, two).

We also define the operator $\hat h(t)=\sum_{j=1}^{N_a} \hat \phi_j(t)e^{-i\delta t}\sigma_{j}^{\dagger}$ so that, from the above Hamiltonian (Eq.~(\ref{ham})), the equations of motion have the compact form of 
\begin{subequations}
\begin{align}
\ket{\dot \psi_{2,0}} = -ig&  \hat h^{\dagger}(t) \ket{\psi_{1,1}(t)}   \label{11a} \\
\ket{\dot \psi_{1,1}} =   -ig&  \hat h(t)\ket{\psi_{2,0}(t)} -ig  \hat h^{\dagger}(t) \ket{\psi_{0,2}(t)} \label{11b} \\
\ket{\dot \psi_{0,2}} = -ig&  \hat h(t)\ket{\psi_{1,1}(t)} \label{11c}
\end{align} 
\end{subequations}

We start by formally integrating Eq.~(\ref{11a}) and substituting it into Eq.~(\ref{11b}), then normal-ordering the $\hat h(t)\hat h^{\dagger}(t_1)$ term under the integral sign, using the commutator (\ref{phicom2}).  The result is  
\begin{equation}
\ket{\dot \psi_{1,1}}=-g^2\sum_{j,k=1}^{N_a}\theta_{j,k} \sigma_{j}^{\dagger}\sigma_{k}\ket{\psi_{1,1}(t)}-ig\hat h(t)\ket G \ket{\psi_I}-ig\hat h^{\dagger}(t)\ket{\psi_{0,2}(t)}-g^2\int_{-\infty}^{t}d t_1 \sum_{j,k=1}^{N_a} \hat \phi_{k}^{\dagger}(t_1)  \hat \phi_j(t)  \sigma^{\dagger}_j \sigma_k e^{-i\delta (t-t_1)}\ket{\psi_{1,1}(t_1)}
\label{e511}
\end{equation}
where $\ket G$ is the initial atomic state (with all atoms in the ground state), and $\ket{\psi_I}$ is the initial photon state. We then define the operator  $\hat \Gamma=g^2\sum_{j,k=1}^{N_a}\theta_{j,k} \sigma_{j}^{\dagger}\sigma_{k}$, which we use to get the formal solution
\begin{align}
\ket{\psi_{1,1}}=&-ig\int_{-\infty}^{t}d t_1e^{-\hat \Gamma(t-t_1)} \hat h(t_1)\ket G \ket{\psi_I}-ig \int_{-\infty}^{t}d t_1e^{-\hat \Gamma(t-t_1)} h^{\dagger}(t_1)\ket{\psi_{0,2}(t_1)}\cr
&\quad-g^2\int_{-\infty}^{t}d t_1e^{-\hat \Gamma(t-t_1)} \int_{-\infty}^{t_1}dt_2 \sum_{j,k=1}^{N_a} \hat \phi_{k}^{\dagger}(t_2)  \hat \phi_j(t_1)  \sigma^{\dagger}_j \sigma_k e^{-i\delta (t_1-t_2)}\ket{\psi_{1,1}(t_2)}
\label{psi11}
\end{align}

We now substitute $\ket{\psi_{1,1}(t)}$ into itself, and again put the field operators in normal order, making use of the result established in the Appendix of \cite{us}, that in the Markovian approximation all field operators whose time variables, $t_i$ and $t_j$, are nested more than 1 level apart (for instance, $t_1$ and $t_3$, when $t_1>t_2>t_3$) effectively commute.  Then only the term containing the initial field state survives when substituted in the last term of (\ref{psi11}), with the result
\begin{align}
\ket{\psi_{1,1}(t)}=&-ig\int_{-\infty}^{t}d t_1e^{-\hat \Gamma(t-t_1)} \hat h(t_1)\ket G \ket{\psi_I}-ig \int_{-\infty}^{t}d t_1e^{-\hat \Gamma(t-t_1)} \hat h^{\dagger}(t_1) \ket{\psi_{0,2}(t_1)}\cr
&\quad+ig^3\int_{-\infty}^{t}d t_1 \int_{-\infty}^{t_1}dt_2 \int_{-\infty}^{t_2}d t_3 e^{-\hat \Gamma(t-t_1)}\sum_{j,k=1}^{N_a} \hat \phi_{k}^{\dagger}(t_2)  \hat \phi_j(t_1)  \sigma^{\dagger}_j \sigma_k e^{-i\delta (t_1-t_2)}e^{-\hat \Gamma(t_2-t_3)} \hat h(t_3)\ket G \ket{\psi_I}
\label{psi11f}
\end{align}

As this is in terms of just the initial state and $\ket{\psi_{0,2}(t)}$ we can substitute it into Eq.~(\ref{11c}). Doing so will kill the triple integral term, as $\hat h(t)$ contains photon lowering operators that will commute with $\hat \phi^{\dagger}(t_2)$. The differential equation for the state when both photons have been absorbed is then
\begin{equation}
\ket{\dot \psi_{0,2}} = -g^2\int_{-\infty}^{t}d t_1 \hat h(t)e^{-\hat \Gamma(t-t_1)} \hat h(t)\ket G \ket{\psi_I} -g^2\int_{-\infty}^{t}d t_1 \hat h(t) e^{-\hat \Gamma(t-t_1)}\hat h^{\dagger}(t_1) \ket{\psi_{0,2}(t_1)}
\label{eq515}
 \end{equation}
 
The photon operators in the last term in (\ref{eq515}) can be brought into normal order past the exponential of the operator $\hat\Gamma$, which only contains atomic operators.  The commutator then brings a delta function of $t-t_1$ which causes the integral to vanish.  The final expression reduces to 
\begin{equation}
 \ket{\dot \psi_{0,2}} = -\hat \Gamma \ket{\psi_{0,2}(t)}-g^2\int_{-\infty}^{t}d t_1 \hat h(t) e^{-\hat \Gamma(t-t_1)} \hat h(t)\ket G \ket{\psi_I}
 \end{equation}
which integrates to
\begin{equation}
\ket{\psi_{0,2}(t)}=-g^2\int_{-\infty}^{t}d t_1\int_{-\infty}^{t_1}d t_2 e^{-\hat \Gamma(t-t_1)} \hat h(t_1)e^{-\hat \Gamma(t_1-t_2)} \hat h(t_2)\ket G \ket{\psi_I}
\label{psi02f}
\end{equation}

Finally, we can substitute this into (\ref{psi11f}), and the result into the equation (\ref{11a}) for $\ket{\psi_{2,0}(t)}$, which can then also be integrated directly.  Multiplying both sides by $\bra G$  yields (for sufficiently large $t$) the final two-photon state
\begin{align}
|\psi_{g}(t)\rangle=&|\psi_I\rangle -g^2\int_{-\infty}^{t}dt_1\int_{-\infty}^{t_1}d t_2\bra G\hat h^{\dagger}(t_1) e^{-\hat \Gamma(t_1-t_2)} \hat h(t_2)\ket G \ket{\psi_I}\cr
&+g^4\int_{-\infty}^{t}d t_1 \int_{-\infty}^{t_1}dt_2 \int_{-\infty}^{t_2}d t_3 \int_{-\infty}^{t_2}d t_4 \Bigg[\bra G\hat h^{\dagger}(t_1) e^{-\hat \Gamma(t_1-t_2)}\sum_{j,k=1}^{N_a} \hat\phi_{k}^{\dagger}(t_3)\hat \phi_j(t_2) \sigma^{\dagger}_j \sigma_k 
e^{-\hat \Gamma(t_3-t_4)} \hat h(t_4)\ket G \ket{\psi_I}\cr
&\qquad\qquad+\bra G \hat h^{\dagger}(t_1) e^{-\hat \Gamma(t_1-t_2)}\hat h^{\dagger}(t_2)e^{-\hat \Gamma (t_2-t_3)}\hat h(t_3) e^{-\hat \Gamma(t_3-t_4)} \hat h(t_4)\ket G \ket{\psi_I}\Bigg]
   \label{main}
\end{align}
While complicated by the presence of the atomic operators in the exponentials, this clearly has the same structure of the solution for  two atoms that we presented in Eq.~(31) of \cite{us2}).  The ``nonlinear'' terms are the two quadruple integrals.  The first one involves sequential excitation of two atoms, whereas the second one involves two atomic excitations overlapping in time (what we call the ``doubly excited'' state).

It is not very hard to deal formally with the exponential of the atomic operator $\hat\Gamma$.  Inspection shows that, in the first of the quadruple integrals in (\ref{main}), all that is required is to diagonalize the matrix that represents $\hat\Gamma$ in the subspace of states with only one atom excited, which has dimension ${N_a}$ and whose elements can be simply written as $\ket i$, $i=1,\ldots {N_a}$.   We have then
\begin{equation}
    M_{i,j}\equiv \bra i\hat\Gamma\ket j= \bra{i} \sum_{m,n=1}\theta_{m,n}\sigma^{\dagger}_{m}\sigma_{n}\ket{j} =g^2 \theta_{i,j}
\end{equation}
Diagonalizing this as $M=P D P^{-1}$ we can write
\begin{equation}
e^{-\hat \Gamma(t_i-t_j)}= P e^{-g^2 D_1(t_i-t_j)}P^{-1} 
\end{equation}
in the subspace in question.  

For the doubly-excited term, we note that the first $\hat h$ operator ($\hat h(t_4)$) takes us to the subspace with one excitation, where we can use the result we have just derived for the term $e^{-\hat \Gamma(t_i-t_j)}$.  We can perform a similar process for the $e^{-\hat \Gamma (t_3-t_4)}$, but then the next operator, $\hat h(t_3)$, takes us to the subspace with two excitations, so we need the form of $e^{-\hat \Gamma (t_2-t_3)}$ in that subspace.  If we label the two excited atoms $l$ and $m$, with $l<m$ to avoid double counting, we see that this subspace has dimension ${N_a}({N_a}-1)/2$, and that
\begin{equation}
    \bra{ l}\bra{ m} \hat \Gamma \ket{ r}\ket{ s}=\theta_{ l, s}\delta_{ m, r}+\theta_{ l, r}\delta_{ m, s} +\theta_{ m, s}\delta_{ l, r}+\theta_{ m, r}\delta_{ l, s}
\end{equation}
At this point, some bookkeeping is necessary.  For any ${N_a}$, we create a list of pairs $(l,m)$, with $1\le l< {N_a}$ and $l,m\le {N_a}$, and number the elements by a single index $j$, $1\le j\le {N_a}({N_a}-1)/2$.  We call $l\equiv j[1]$ and $m\equiv j[2]$ the values of $l$ and $m$ corresponding to the state $j$.  We can then introduce a matrix $M^\prime$ with elements $M^\prime_{j,k} = \bra{j[1]}\bra{j[2]} \hat\Gamma \ket{k[1]}\ket{k[2]}$, and its diagonal representation $D^\prime$, satisfying  $M^\prime =Q D^\prime Q^{-1}$, which allows us to write 
\begin{equation}
    e^{-\hat \Gamma(t_i-t_j)}= Q e^{-g^2 D^\prime(t_i-t_j)}Q^{-1} 
\end{equation}
in the two-excitation subspace.  All this allows us to write Eq.~(\ref{main}) entirely in terms of photon operators and matrix elements. We assume that the initial state is of the form 
\begin{equation}
    \ket{\psi_I}=\int d t_1 dt_2 f(t_1) f(t_2) \hat A^{\dagger}(t_1)\hat B^{\dagger}(t_2)\ket{0}
\end{equation}
(i.e., two counterpropagating photons with identical envelopes), and define the following sums: 
\begin{equation}
    \Sigma_{i}^{\pm_a,\pm_b}=\sum_{ j, k=1}^{{N_a}}P_{j,i}P^{-1}_{i,k}e^{ik_F(\pm_a z_j \pm_b z_k)}
\end{equation}
\begin{align}
    \chi_{p,q,r}^{\pm_a,\pm_b}= &\sum_{k,l=1}^{{N_a}({N_a}-1)/2}\sum_{j,m=1}^{N_a} Q_{k,r}Q^{-1}_{r,l}\hat e^{\pm_a ik_F z_j} \Big(P_{j,p}P^{-1}_{p,k[1]} e^{\pm_b ik_F z_{k[2]}}+P_{j,p}P^{-1}_{p,k[2]}e^{\pm_b ik_F z_{k[1]}}\Big)\cr
    &\quad\times \Big(P_{l[1],q}P^{-1}_{q,m}\big(e^{i k_F(-z_{l[2]}+z_{m}) }+e^{i k_F(z_{l[2]}-z_{m})}\big)+P_{l[2],q}P^{-1}_{q,m}\big(e^{i k_F(-z_{l[1]}+z_{m}) }+e^{i k_F(z_{l[1]}-z_{m})}\big)\Big)
\end{align}
With this, the final scattered spacetime envelope for the pulses, corresponding to both counterpropagating photons passing through, is 
\begin{align}
    f_{a,b}(t_1,t_2)=f(t_1)f(t_2)-g^2\sum_{i=1}^{{N_a}}\Big( \Sigma_{i}^{-,+} G_{\Gamma_i}(t_1)f(t_2)+ \Sigma_{i}^{+,-} G_{\Gamma_i}(t_2)f(t_1)\Big)\cr
    +g^4\sum_{p,q=1}^{{N_a}} \Big(\Sigma_{p}^{-,+}\Sigma_{q}^{+,-}+\Sigma_{p}^{-,-}\Sigma_{q}^{+,+}\Big)\Big[\theta(t_1-t_2) G_{\Gamma_p}(t_1)G_{\Gamma_q}(t_2)+\theta(t_2-t_1) G_{\Gamma_p}(t_2)G_{\Gamma_q}(t_1)\Big]\cr
    -g^4\sum_{p,q=1}^{{N_a}} \Big(\Sigma_{p}^{-,+}\Sigma_{q}^{+,-}+\Sigma_{p}^{-,-}\Sigma_{q}^{+,+}\Big) \Big[\theta(t_1-t_2)e^{-\Gamma_p(t_1-t_2)}G_{\Gamma_p}(t_2)G_{\Gamma_q}(t_2)+\theta(t_2-t_1) e^{-\Gamma_p(t_2-t_1)} G_{\Gamma_p}(t_1)G_{\Gamma_q}(t_1)  \Big]\cr
    +g^4 \sum_{p,q=1}^{N_a} \sum_{r=1}^{{N_a}({N_a}-1)/2} \Big[\chi_{p,q,r}^{-,+}\theta(t_1-t_2) e^{-\Gamma_p(t_1-t_2)} \mathcal{E}_{r,q}(t_2)+\chi_{p,q,r}^{+,-}\theta(t_2-t_1) e^{-\Gamma_p(t_2-t_1)} \mathcal{E}_{r,q}(t_1)\Big]
    \label{manySplit}
\end{align}
We have labeled the eigenvalues of $D$ as $\Gamma_p$, and those of $D^\prime$ as $\Gamma^\prime_r$, in terms of which the functions $G_\Gamma$ and $\mathcal{E}_{r,p}$ are given by 
\begin{align}
    G_{\Gamma}(t) = e^{-\Gamma t}\int_{-\infty}^{t}dt^{\prime} e^{\Gamma t^{\prime}}f(t^{\prime}) && \mathcal{E}_{r,p}(t) =\int_{-\infty}^{t} d t^{\prime}e^{-\Gamma^\prime_r(t- t^{\prime})}  f(t^{\prime})G_{\Gamma_p}(t^{\prime})
\label{e34}
\end{align}

Unfortunately, for a Gaussian pulse it is impossible to derive an analytic form for $\mathcal{E}_{r,p}$, making a computational solution in the time domain prohibitively expensive. Fortunately, by translating these into the frequency domain it becomes possible to create efficient code that can calculate factors like the norm and fidelity. 

It is trivial to translate terms that are separable in $t_1$ and $t_2$ to the frequency domain. Defining $\tilde f(\omega)=\frac{1}{\sqrt{2\pi}}\int d t e^{i\omega t} f(t)$ as the Fourier transform of $f$, it turns out that $\tilde G_{\Gamma}(\omega)=\tilde f(\omega)/(\Gamma-i\omega)$. Additionally, it is relatively straightforward to show that the double Fourier transforms of the entangled terms are 
\begin{align}
    \frac{1}{2\pi}\int d t_1 dt_2 e^{i\omega_1 t_1} e^{i \omega_2 t_2} \Big(\theta(t_1-t_2)e^{-\Gamma_p(t_1-t_2)}
G_{\Gamma_p}(t_2)G_{\Gamma_q}(t_2)+\theta(t_2-t_1)e^{-\Gamma_p(t_2-t_1)}
G_{\Gamma_p}(t_1)G_{\Gamma_q}(t_1)\Big)=\cr
\Bigg(\frac{1}{\Gamma_p-i\omega_1}+\frac{1}{\Gamma_p-i\omega_2}\Bigg) \frac{\tilde f_{\text{ent},\Gamma_p}(\omega_1+\omega_2)+ \tilde f_{\text{ent},\Gamma_q}(\omega_1+\omega_2) } {\Gamma_p+\Gamma_q-i(\omega_1+\omega_2)}
\end{align}
\begin{equation}
    \frac{1}{2\pi}\int d t_1 dt_2 e^{i\omega_1 t_1} e^{i \omega_2 t_2} \theta(t_1-t_2)e^{-\Gamma_p(t_1-t_2)}
\mathcal{E}_{r,q}(t_2)=
 \frac{\tilde f_{\text{ent},\Gamma_q}(\omega_1+\omega_2)} {(\Gamma_p-i\omega_1)(\Gamma^\prime_r-i(\omega_1+\omega_2))}
\end{equation}
In these equations, 
\begin{equation}
    \tilde f_{\text{ent},\Gamma_q}(\omega_1+\omega_2)=\int d\omega_b \frac{\tilde f(\omega_1+\omega_2-\omega_b)\tilde f(\omega_b)}{\Gamma_q-i\omega_b}
\end{equation}
which is very similar in form to Eq.~(A7) in \cite{us} and, for a Gaussian pulse of $\tilde f(\omega)=\frac{e^{-\omega^2/4\sigma_\omega}}{\sqrt{\sigma_\omega \sqrt{2\pi}}}$, is
\begin{equation}
\tilde f_{\text{ent},\Gamma_q}(\omega_1+\omega_2)=\frac{e^{-((i+1)\Gamma_q+\omega)((i-1)\Gamma_q+\omega)/4\sigma_\omega^2}}{2\sqrt{2\pi}\sigma_\omega}\text{erfc} \Bigg(\frac{2 \Gamma_q-i\omega}{2\sqrt{2}\sigma_\omega}\Bigg)
\end{equation}
Substituting these equations into Eq.~(\ref{manySplit}) the final, two-photon scattered spectrum for two initially unentangled, counter-propagating, identical photons through ${N_a}$ arbitrarily spaced emitters is
\begin{align}
    \tilde f_{a,b}(\omega_1,\omega_2)=\tilde f(\omega_1) \tilde f(\omega_2) \Bigg[\Bigg(1-g^2\sum_{i=1}^{{N_a}} \frac{\Sigma_{i}^{-,+}}{\Gamma_i-i\omega_1}\Bigg)\Bigg(1-g^2\sum_{i=1}^{{N_a}}\frac{\Sigma_{i}^{+,-}}{\Gamma_i-i\omega_2}\Bigg) +g^4\sum_{p,q=1}^{{N_a}} \frac{\Sigma_{p}^{-,-}\Sigma_{q}^{+,+}}{(\Gamma_p-i\omega_1)(\Gamma_q-i\omega_2)}\Bigg]\cr
    -g^4\sum_{p,q=1}^{{N_a}} \Big(\Sigma_{p}^{-,+}\Sigma_{q}^{+,-}+\Sigma_{p}^{-,-}\Sigma_{q}^{+,+}\Big)\Bigg(\frac{1}{\Gamma_p-i\omega_1}+\frac{1}{\Gamma_p-i\omega_2}\Bigg) \frac{\tilde f_{\text{ent},\Gamma_p}(\omega_1+\omega_2)+ \tilde f_{\text{ent},\Gamma_q}(\omega_1+\omega_2) } {\Gamma_p+\Gamma_q-i(\omega_1+\omega_2)}\cr
    +g^4 \sum_{p,q=1}^{N_a} \sum_{r=1}^{{N_a}({N_a}-1)/2} \Big[\frac{\chi_{p,q,r}^{-,+}}{\Gamma_p-i\omega_1} +\frac{\chi_{p,q,r}^{+,-}}{\Gamma_p-i\omega_2}\Big]\frac{\tilde f_{\text{ent},\Gamma_q}(\omega_1+\omega_2)} {\Gamma^\prime_r-i(\omega_1+\omega_2)}
    \label{manySplitfreq}
\end{align}
where we have made use of the fact that, due to symmetry, $\Sigma_{p}^{-,+}\Sigma_{q}^{+,-}+\Sigma_{p}^{-,-}\Sigma_{q}^{+,+}=\Sigma_{q}^{-,+}\Sigma_{p}^{+,-}+\Sigma_{q}^{-,-}\Sigma_{p}^{+,+}$. 

Finally, from the structure of this equation we can essentially read off the single photon transmission coefficients for a photon transmitting from the right $t_a(\omega)$ and from the left, $t_b(\omega)$, as the first line of Eq.~(\ref{manySplitfreq}) contains a term that looks like two independent transmission events. We note that we have indeed rigorously derived these quantities and that they are consistent with transmission coefficients derived using the meth od in \cite{zubairy1} and applying the Markovian approximation.
\begin{align}
    t_a(\omega)=1-g^2\sum_{i=1}^{{N_a}} \frac{\Sigma_{i}^{-,+}}{\Gamma_i-i\omega} && t_b(\omega)=1-g^2\sum_{i=1}^{{N_a}} \frac{\Sigma_{i}^{+,-}}{\Gamma_i-i\omega}
\end{align}

By symmetry, we clearly have $t_a = t_b \equiv t_N$ (the single-photon transmission coefficient  for $N=N_a/2$ sites mentioned in the main text).  In calculating the fidelity in Fig. 4 and 5 of the main text we used the following equation.

\begin{equation}
    \sqrt{\mathcal{F}}e^{i\Phi}=\int d\omega_1 d\omega_2 t_a(\omega_1)^* t_b(\omega_2)^* \tilde f(\omega_1)^* \tilde f(\omega_2)^* \tilde f_{a,b}(\omega_1,\omega_2)
\end{equation}

\section{Analytic Approximation For Non-Interacting Atoms}
The result (\ref{manySplitfreq}) in the previous section is exact within the Markovian approximation.  It is also amenable to numerical evaluation, at least for reasonable numbers of atoms ${N_a}$, and this is how we have obtained all the numerical results in the text.  However, it is not very transparent, so for insights on what is going on in the system we present in this section an approximate solution, based on the concatenation of the single-site scattering matrix (where a single site is a pair of TLEs) $N$ times over, along the lines of the paper \cite{brod2}. 

We recall here that the scattering matrix connects the scattered spectrum, $\tilde f_{ab}(\omega_1,\omega_2)$ to the incoming spectrum, here assumed to be of the form $\tilde f(\omega_a)\tilde f(\omega_b)$, through the equation
\begin{equation}
\tilde f_{ab}(\omega_a,\omega_b) = \int d\nu_a d\nu_b\,S(\omega_1,\omega_2,\nu_a,\nu_b) \tilde f(\nu_a) \tilde f(\nu_b).
\end{equation}

As mentioned in the text, under near-unit transmission conditions, we find, from the results in \cite{us2},
the following approximate two-photon scattering matrix (for counterpropagating photons), for a single atom pair: 
\begin{align}
   & S_1(\omega_1,\omega_2,\nu_a,\nu_b)= t(\omega_1)\delta(\omega_1-\nu_a) t(\omega_2)\delta(\omega_2-\nu_b)-\frac{\cos^2\phi}{\pi} e^{-2i\phi} \delta(\omega_1+\omega_2-\nu_a-\nu_b) \cr   
    &\times\biggl[\Gamma_+^2\left(\frac{1}{\Gamma_+ -i\omega_1}+ \frac{1}{\Gamma_+ -i\omega_2}\right) \frac{1}{(\Gamma_+ -i\nu_a)(\Gamma_+ -i\nu_b)} +\Gamma_-^2\left(\frac{1}{\Gamma_- -i\omega_1}+ \frac{1}{\Gamma_- -i\omega_2}\right) \frac{1}{(\Gamma_- -i\nu_a)(\Gamma_- -i\nu_b)} \biggr] 
\label{jgb1}
\end{align}
where 
\begin{equation}
\Gamma_{\pm} = \frac{\Gamma e^{i\phi}}{\cos\phi}(1\pm\cos\phi)
\label{jgb2}
\end{equation}
(with $\Gamma \equiv g^2$) and 
\begin{equation}
t(\omega) = -\exp[-2i\phi-2i\omega\Gamma/\delta^2]
\label{jgb3}
\end{equation}

Equation (\ref{jgb1}) can be derived from the Fourier transform of Eq.~(45) in \cite{us2} for the function $f_{ab}$, with a few approximations, some of which are mentioned immediately below Eq.~(48) in \cite{us}.  One is to neglect the contribution from the doubly-excited state; the other, to neglect the product of two reflection coefficients $\rho(t_1)\rho(t_2)$.  We do want to keep the dependence on $\omega$ to lowest order in the transmission coefficient, however.  This is already given in the frequency domain by Eq.~(16) in \cite{us}.  We need to set $\phi$ as a constant (independent of frequency) in that equation (the ``Markovian approximation''), then let $\delta = -\Gamma \tan\phi$, and then expand the logarithm of the prefactor in a power series of $x\equiv\omega/\Gamma$:
\begin{equation}
\ln\left[\frac{(x-\tan\phi)^2}{e^{2i\phi}-(1-i(x-\tan\phi))^2}\right] \simeq i\pi-2i\phi + 2ix\cot^2\phi
\end{equation}
This gives the approximate transmission coefficient (\ref{jgb3}) (making use of the condition $\tan\phi = -\delta/\Gamma$).   

Using this result we can write the $N$-site scattering matrix as 
\begin{align}
    &S_1(\omega_1,\omega_2,\nu_a,\nu_b)= t^N(\omega_1)\delta(\omega_1-\nu_a) t^N(\omega_2)\delta(\omega_2-\nu_b)-\frac{\cos^2\phi}{\pi} e^{-2i\phi} \delta(\omega_1+\omega_2-\nu_a-\nu_b)\cr   
    &\quad\times\biggl[\Gamma_+^2\left(\frac{1}{\Gamma_+ -i\omega_1}+ \frac{1}{\Gamma_+ -i\omega_2}\right) \frac{1}{(\Gamma_+ -i\nu_a)(\Gamma_+ -i\nu_b)}+\Gamma_-^2\left(\frac{1}{\Gamma_- -i\omega_1}+ \frac{1}{\Gamma_- -i\omega_2}\right) \frac{1}{(\Gamma_- -i\nu_a)(\Gamma_- -i\nu_b)} \biggr] \cr
& \quad\times e^{-4(N-1)i\phi}\sum_{j=1}^N \left(e^{2i(\omega_1+\nu_a)\Gamma/\delta^2}\right)^{N-j} \left(e^{2i(\omega_2+\nu_b)\Gamma/\delta^2}\right)^{j-1}
\end{align}
As shown in \cite{brod2}, the sum over $j$ is the essential ingredient to get rid of spectral entanglement.  Following the same steps as in that paper (Eqs.~(60) through (64)) we obtain, in the large $N$ limit, the result
\begin{equation}
e^{-4(N-1)i\phi}\sum_{j=1}^N \left(e^{2i(\omega_1+\nu_a)\Gamma/\delta^2}\right)^{N-j} \left(e^{2i(\omega_2+\nu_b)\Gamma/\delta^2}\right)^{j-1}\simeq \frac{\pi\delta^2}{\Gamma}  t^{N-1}(\omega_1)t^{N-1}(\omega_2)\delta(\omega_1-\omega_2-\nu_a+\nu_b)
\end{equation}
The product of delta functions can then be handled as in Eq.~(65) of [2].  The final result is still entangled, but becomes a product state in the strong adiabatic limit, when we let $\Gamma_\pm \gg \sigma_\omega$.  One finds then
\begin{equation}
S_1(\omega_1,\omega_2,\nu_a,\nu_b)\simeq t^N(\omega_1)\delta(\omega_1-\nu_a) t^N(\omega_2)\delta(\omega_2-\nu_b)\left[1-\frac{\delta^2}{\Gamma}\cos^2\phi \,e^{2i\phi}\left(\frac{1}{\Gamma_+}+\frac{1}{\Gamma_-}\right)\right]
\end{equation}
Using $\tan\phi = -\delta/\Gamma$ and the definition of $\Gamma_\pm$, it is straightforward to verify that the term in square brackets is simply equal to $-e^{2i\phi}$.  Hence, the two-photon wavefunction is equal to the product of two single-photon ones, times the phase factor $-e^{2i\phi}=e^{i(2\phi\pm\pi)}$.  

Unfortunately, it is impossible to make the ``usable phase'' $\Phi=2\phi\pm\pi$ equal to $\pi$ or $-\pi$, since that requires $\phi=0$ or $2\pi$, and hence $\delta =0$.  It is clear from the results in \cite{us}, however, that the single-photon transmission window for non-interacting atoms vanishes as $\delta\to 0$.  Hence the compromise solution we have adopted, which is to set $\delta = \Gamma$, $\phi=-\pi/4$, and thus $\Phi = \pi/2$.  This is also a good choice from the point of view of trying to reach the adiabatic limit, which requires  both $\Gamma_+$ and $\Gamma_-$ to be large; for $\phi = -\pi/4$, one has $|\Gamma_\pm| = \Gamma(\sqrt2 \pm 1)$, which means that, although $\Gamma_-$ is smaller than $\Gamma$, it only takes an extra factor of 2 to reach the $\Gamma_- \gg \sigma_\omega$ regime.

\section{Details on the optimization of the non-interacting array of atoms}

In order to derive the optimized spacing for the non-interacting array of $N_a$ atoms that leads to high transmission, we start with the solution for a single pair presented in \cite{us2} and treat the system as an array of $N=N_a/2$ pairs, as shown in Figure \ref{sys2} below.  

In the Markovian approximation, the phase shift experienced by a photon when traveling from location $z_i$ to location $z_j$ is written simply in terms of the central wavevector as $k_f(z_j-z_i)$, neglecting the variation of $k$ over the components that make up the wavepacket.  Thus, if the distance between two adjacent pairs is $a$, as  in the figure, we treat this as a constant (frequency-independent) multiplicative factor of  $e^{i\phi_a} = e^{ik_F a}$ in all the calculations that follow.

\begin{figure}[H]
\centering
\centerline{\includegraphics[width=.6\columnwidth]{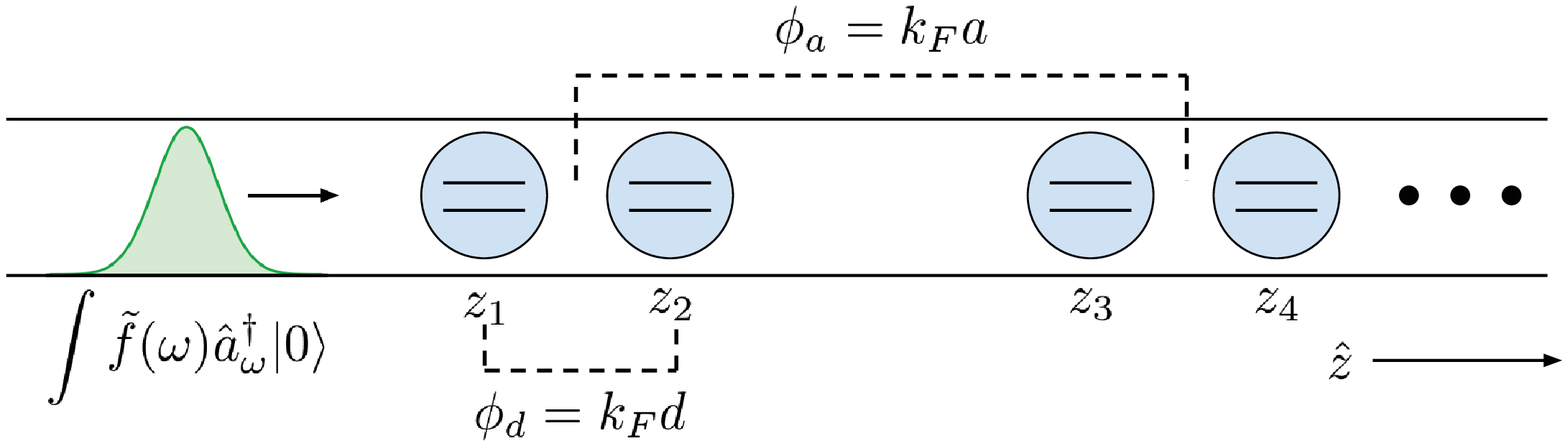}}
\caption{   A diagram of the system of pairs of atoms with the pair distances labeled with $\phi_d$ corresponding to the distance between two atoms in a pair and $\phi_a$ between successive pairs.}
\label{sys2}
\end{figure}

The single-pair transmission coefficient, on the other hand, varies over the wavepacket's frequency components in a non-negligible way. The approximation (\ref{jgb3}) above is fine for a single site, but over $N$ sites, if $N$ is large, one needs to keep the next-order contributions in $\omega/\Gamma$, which result in $|t(\omega)| <1$. Note that in the main text we argued that, for the gate to work, we need $\sigma_\omega d/c \ll \sigma_\omega/\Gamma$, so our concern with the cumulative effect of terms of order $\sigma_\omega/\Gamma$ is still compatible with the Markovian approximation.

To maximize the transmission through the array, then, we describe a pair of TLEs by the full single-photon, two-atom transmission and reflection coefficients we derived in Eq.~(16) and (17) of \cite{us2}, and optimize the distance $d$ between it and the next pair for maximum transmission by treating the two pairs as an ``optical cavity.'' We then find the overall reflection and transmission coefficients of the ``cavity'' by summing up a series of reflection and transmission events, just as we did for the single pair in Eqs. (18)--(22) of \cite{us2}. 

In this way we were able to find a complicated, but still analytic, expression for the transmission of the two-pair system as a function of $\omega$ and $\phi_a$, which is maximum, at $\omega=0$ (which corresponds to the wavepacket's central frequency), when $\phi_a=\pi-\arctan\left[(\Gamma^2-\delta^2)/(2\Gamma \delta)\right]$. For the choice of $\delta/\Gamma=1$ this gives an optimal spacing of $\phi_a=\pi$.

We then repeated this process several more times; we used the analytic transmission coefficient for two pairs separated by a phase of $\pi$ and considered a ``cavity'' of four pairs (eight atoms) with the center of each array of two pairs separated by a new, undefined phase difference $\phi_a$. We optimized this system in the same way, finding where the reflection coefficient was minimized near resonance. Finally we iterated the process to find the optimal phase between two arrays of eight atoms, then two arrays of sixteen atoms. This process led to the optimal spacing see in Fig. 3 of the main text, where the optimal arrangement consists of a unit cell of four atoms separated by $\phi_d$, $\phi_d/3$, and $\phi_d$ in that order. The closest atoms between each successive cell are then also separated by $\phi_d$.

\begin{figure}[h]
\includegraphics[width=.6\columnwidth]{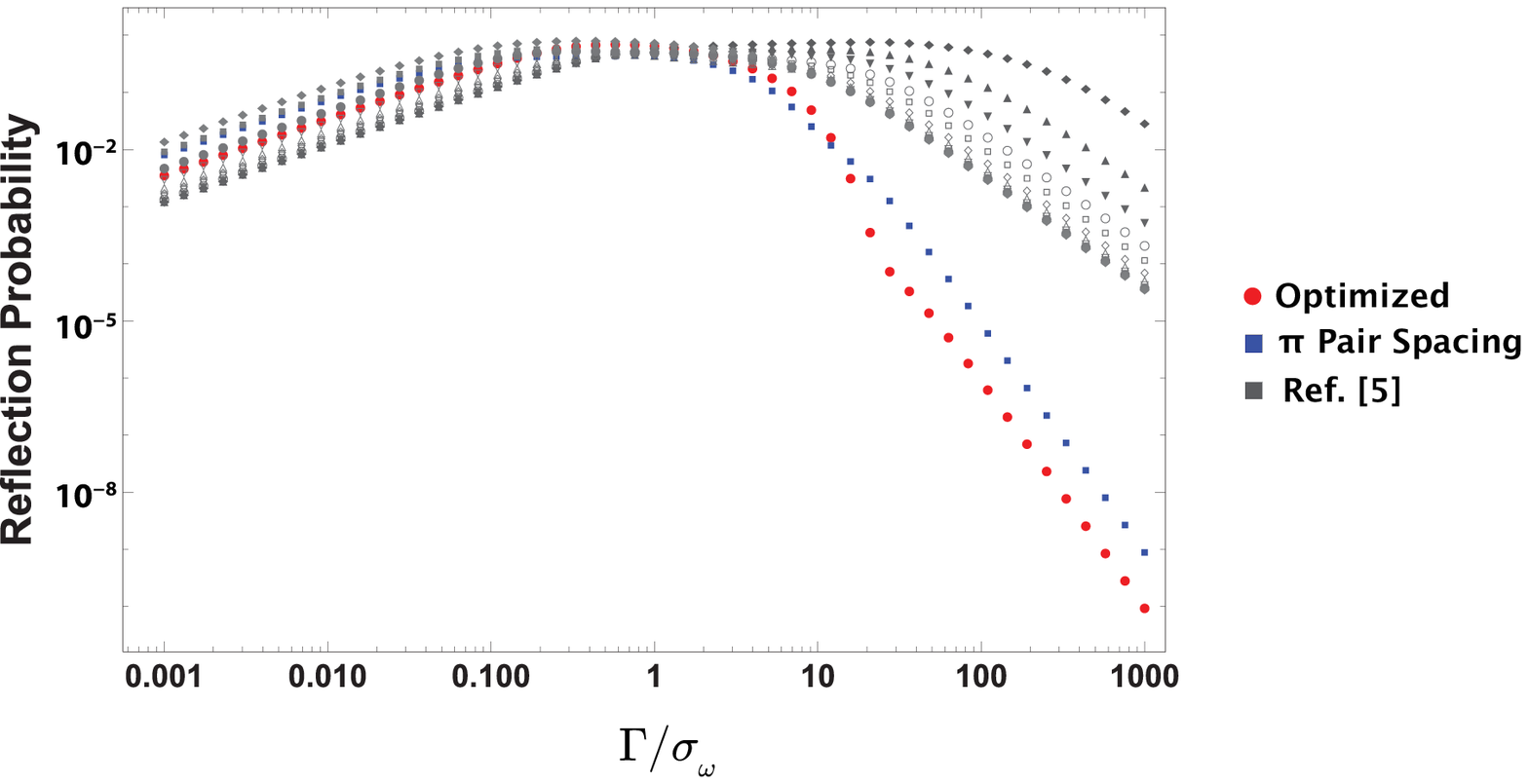}
\caption{Plot of the reflection probability as a function of $\Gamma/\sigma_\omega$ for different inter-atomic distances. The points labeled ``$\pi$ Pair Spacing" correspond to pairs that are evenly spaced at a distance of $\phi_a=\pi$, the ones labeled ``Optimized" correspond to the optimal spacing shown in Fig. 3 of the main text, and the grey curves represent the different transmission maxima given in \cite{law} for an array of equally spaced atoms.}
\label{Romega}
\end{figure}

In Fig. \ref{Romega}, to show that this optimal spacing does indeed work better than other possible choices, we plot the reflection probability for a single photon with a Gaussian profile. We use: the optimal spacing described here, a spacing where the atoms are arranged in pairs and each pair is separated by $\phi_a=\pi$ from its neighbors, and a lattice with the phases between atoms all being the same and given by the maxima found in the  paper \cite{law} by Tsoi and Law. As can be seen, in the adiabatic limit of $\Gamma/\sigma_\omega \gg 1$ (when the gate functions best) our optimal spacing leads to a significantly lower reflection probability.


\end{document}